\documentclass[aps,prl,twocolumn,superscriptaddress,showpacs,preprintnumbers,amsmath,amssymb]{revtex4}
\usepackage{graphicx,color} 
\usepackage{dcolumn}  
\usepackage{subfigure}
\usepackage{multirow}
\RequirePackage{xspace}
\def\Dbar{\kern 0.2em\overline{\kern -0.2em D}{}\xspace}

\def\Bbar    {\kern 0.2em\overline{\kern -0.2em B}{}\xspace}
\def\CP{\ensuremath{C\!P}\xspace}
\graphicspath{{ps}}

\begin{document}
\title{ \quad\\[1.0cm] \textbf{Evidence for the suppressed decay \boldmath$B^{-}\rightarrow DK^{-}, D\rightarrow K^{+}\pi^{-}\pi^{0}$} }

\noaffiliation \affiliation{University of the Basque Country
UPV/EHU, 48080 Bilbao} \affiliation{Beihang University, Beijing
100191} \affiliation{University of Bonn, 53115 Bonn}
\affiliation{Budker Institute of Nuclear Physics SB RAS and
Novosibirsk State University, Novosibirsk 630090}
\affiliation{Faculty of Mathematics and Physics, Charles University,
121 16 Prague}
\affiliation{University of Cincinnati, Cincinnati, Ohio 45221}
\affiliation{Deutsches Elektronen--Synchrotron, 22607 Hamburg}
\affiliation{Department of Physics, Fu Jen Catholic University,
Taipei 24205} \affiliation{Justus-Liebig-Universit\"at Gie\ss{}en,
35392 Gie\ss{}en}
\affiliation{Hanyang University, Seoul 133-791}
\affiliation{University of Hawaii, Honolulu, Hawaii 96822}
\affiliation{High Energy Accelerator Research Organization (KEK),
Tsukuba 305-0801} \affiliation{Hiroshima Institute of Technology,
Hiroshima 731-5193} \affiliation{Ikerbasque, 48011 Bilbao}
\affiliation{Indian Institute of Technology Guwahati, Assam 781039}
\affiliation{Indian Institute of Technology Madras, Chennai 600036}
\affiliation{Institute of High Energy Physics, Vienna 1050}
\affiliation{Institute for High Energy Physics, Protvino 142281}
\affiliation{INFN - Sezione di Torino, 10125 Torino}
\affiliation{Institute for Theoretical and Experimental Physics,
Moscow 117218} \affiliation{J. Stefan Institute, 1000 Ljubljana}
\affiliation{Kanagawa University, Yokohama 221-8686}
\affiliation{Institut f\"ur Experimentelle Kernphysik, Karlsruher
Institut f\"ur Technologie, 76131 Karlsruhe}
\affiliation{Korea Institute of Science and Technology Information,
Daejeon 305-806} \affiliation{Korea University, Seoul 136-713}
\affiliation{Kyungpook National University, Daegu 702-701}
\affiliation{University of Maribor, 2000 Maribor}
\affiliation{Max-Planck-Institut f\"ur Physik, 80805 M\"unchen}
\affiliation{School of Physics, University of Melbourne, Victoria
3010} \affiliation{Moscow Physical Engineering Institute, Moscow
115409} \affiliation{Moscow Institute of Physics and Technology,
Moscow Region 141700} \affiliation{Graduate School of Science,
Nagoya University, Nagoya 464-8602} \affiliation{Kobayashi-Maskawa
Institute, Nagoya University, Nagoya 464-8602}
\affiliation{Nara Women's University, Nara 630-8506}
\affiliation{National Central University, Chung-li 32054}
\affiliation{National United University, Miao Li 36003}
\affiliation{Department of Physics, National Taiwan University,
Taipei 10617} \affiliation{H. Niewodniczanski Institute of Nuclear
Physics, Krakow 31-342} \affiliation{Nippon Dental University,
Niigata 951-8580} \affiliation{Niigata University, Niigata 950-2181}
\affiliation{Osaka City University, Osaka 558-8585}
\affiliation{Pacific Northwest National Laboratory, Richland,
Washington 99352} \affiliation{Panjab University, Chandigarh 160014}
\affiliation{University of Pittsburgh, Pittsburgh, Pennsylvania
15260}
\affiliation{Research Center for Electron Photon Science, Tohoku
University, Sendai 980-8578}
\affiliation{Seoul National University, Seoul 151-742}
\affiliation{Soongsil University, Seoul 156-743}
\affiliation{Sungkyunkwan University, Suwon 440-746}
\affiliation{School of Physics, University of Sydney, NSW 2006}
\affiliation{Tata Institute of Fundamental Research, Mumbai 400005}
\affiliation{Excellence Cluster Universe, Technische Universit\"at
M\"unchen, 85748 Garching} \affiliation{Toho University, Funabashi
274-8510} \affiliation{Tohoku Gakuin University, Tagajo 985-8537}
\affiliation{Tohoku University, Sendai 980-8578}
\affiliation{Department of Physics, University of Tokyo, Tokyo
113-0033} \affiliation{Tokyo Institute of Technology, Tokyo
152-8550} \affiliation{Tokyo Metropolitan University, Tokyo
192-0397} \affiliation{Tokyo University of Agriculture and
Technology, Tokyo 184-8588}
\affiliation{CNP, Virginia Polytechnic Institute and State
University, Blacksburg, Virginia 24061} \affiliation{Wayne State
University, Detroit, Michigan 48202} \affiliation{Yamagata
University, Yamagata 990-8560} \affiliation{Yonsei University, Seoul
120-749}
\author{M.~Nayak}\affiliation{Indian Institute of Technology Madras, Chennai 600036} 
\author{J.~Libby}\affiliation{Indian Institute of Technology Madras, Chennai 600036} 
\author{K.~Trabelsi}\affiliation{High Energy Accelerator Research Organization (KEK), Tsukuba 305-0801} 
  \author{I.~Adachi}\affiliation{High Energy Accelerator Research Organization (KEK), Tsukuba 305-0801} 
  \author{H.~Aihara}\affiliation{Department of Physics, University of Tokyo, Tokyo 113-0033} 
  \author{D.~M.~Asner}\affiliation{Pacific Northwest National Laboratory, Richland, Washington 99352} 
  \author{T.~Aushev}\affiliation{Institute for Theoretical and Experimental Physics, Moscow 117218} 
  \author{A.~M.~Bakich}\affiliation{School of Physics, University of Sydney, NSW 2006} 
  \author{A.~Bala}\affiliation{Panjab University, Chandigarh 160014} 
  \author{P.~Behera}\affiliation{Indian Institute of Technology Madras, Chennai 600036} 
% ...
\collaboration{The Belle Collaboration}
\noaffiliation

\begin{abstract}
We report a study of the suppressed decay $B^{-}\to DK^{-}$, $D\to
K^{+}\pi^{-}\pi^{0}$, where $D$ denotes either a $D^{0}$ or a
$\Dbar^0$  meson. The decay is sensitive to the \CP-violating
parameter $\phi_{3}$. Using a data sample of $772\times 10^6$
$B\Bbar$ pairs collected at the $\Upsilon(4S)$ resonance with the
Belle detector, we measure the ratio of branching fractions of the
above suppressed decay to the favored decay $B^{-}\to DK^{-}$, $D\to
K^{-}\pi^{+}\pi^{0}$. Our result is $R_{DK}$ =
$[1.98\pm0.62(\mathrm{stat.})\pm0.24(\mathrm{syst.})]\times10^{-2}$,
which indicates the first evidence of the signal for this suppressed
decay with a significance of $3.2$ standard deviations. We measure
the direct \CP asymmetry between the suppressed $B^{-}$ and $B^{+}$
decays to be $A_{DK} = 0.41\pm0.30 (\mathrm{stat.})\pm0.05
(\mathrm{syst.})$. We also report measurements for the analogous
quantities $R_{D\pi}$ and $A_{D\pi}$ for the decay $B^{-}\to
D\pi^{-},~D\to K^{+}\pi^{-}\pi^{0}$.
\end{abstract}

\pacs{13.25.Hw, 11.30.Er, 12.15.Hh, 14.40.Nd}
\maketitle
\tighten
{\renewcommand{\thefootnote}{\fnsymbol{footnote}}}
\setcounter{footnote}{0}

Several hadronic weak decays related by the combined
charge-conjugation and parity (\CP) transformations exhibit
different behavior. Such violation of \CP symmetry is described by
the Standard Model of particle physics via an irreducible complex
phase in the $3\times3$ Cabibbo-Kobayashi-Maskawa (CKM) quark mixing
matrix \cite{ckm}, which has elements $V_{qq^{\prime}}$, with
$q=u,c,t$ and $q^{\prime}=d,s,b$. The unitarity triangle (UT) is
used to represent the amount of \CP violation parameterized by the
CKM matrix. The UT angle $\phi_3 = \gamma
\equiv\arg{(-V_{ud}V_{ub}^{*}/V_{cd}V_{cb}^{*})}$ is less precisely
measured compared to the other two angles $\phi_{1}(\equiv\beta)$
and $\phi_2(\equiv\alpha)$. The particular importance of improving
the determination of $\phi_3$ lies in the fact that it is the only
\CP-violating parameter that describes the UT that can be measured
solely in tree-level processes. As a result, such measurements
provide a benchmark to search for new physics contributions in
loop-dominated processes that would otherwise constrain the UT.

Various methods to determine $\phi_3$ in the tree decay $B^{-}\to
DK^{-}$, where $D$ is a $D^0$ or $\Dbar^0$ decaying to a common
final state \cite{icc}, have been proposed \cite{glw,soni,giri}. In
this paper, we focus on the Atwood-Dunietz-Soni (ADS) method
\cite{soni} using the decay $B^{-}\to DK^{-}$ followed by $D\to
K^{+}\pi^{-}\pi^{0}$. Several ADS measurements have been made using
$D\to K^{+}\pi^{-}$ \cite{BABARADS,BelleADS,CDFADS,LHCbADS,
Negishi}. However, given a significantly larger branching fraction
for $\Dbar^0\to K^+\pi^{-}\pi^{0}$ $[(13.9\pm0.5)\%]$ than
$\Dbar^0\to K^+\pi^{-}$ $[(3.89\pm0.05)\%]$ \cite{pdg}, the former
mode is potentially more sensitive to $\phi_3$ despite a reduced
acceptance owing to the presence of a $\pi^0$ meson in the final
state. Herein, we search for $B^{-}\rightarrow
[K^{+}\pi^{-}\pi^{0}]_{D}K^{-}$ events for the first time in Belle,
where the favored $B^{-}\rightarrow D^0K^{-}$ decay followed by the
doubly Cabibbo-suppressed (DCS) $D^0\rightarrow K^{+}\pi^{-}\pi^{0}$
decay interferes with the suppressed $B^{-}\rightarrow \Dbar^0K^{-}$
decay followed by the Cabibbo-favored (CF) $\Dbar^0\rightarrow
K^{+}\pi^{-}\pi^{0}$ decay. The interference between the two
amplitudes can lead to a large direct \CP asymmetry between the
suppressed decays. We use $B^{-}\rightarrow D\pi^{-}$ as a control
channel because of the kinematic similarity to $B^{-}\rightarrow
DK^{-}$ and its much larger branching fraction.

One observable measured is the ratio of the suppressed to favored
branching fractions
\begin{eqnarray}
{ R_{DK} } &=&
{\frac{\mathcal{B}([K^{+}\pi^{-}\pi^{0}]_{D}K^{-})+\mathcal{B}(
[K^{-}\pi^{+}\pi^{0}]_{D}K^{+})}{\mathcal{B}(
[K^{-}\pi^{+}\pi^{0}]_{D}K^{-})+\mathcal{B}(
[K^{+}\pi^{-}\pi^{0}]_{D}K^{+})}}\\ \nonumber
&=&r_{B}^{2}+r_{D}^{2}+2r_{B}r_{D}R_{K\pi \pi^{0}}\cos \phi_{3} \cos
(\delta_{B}+ \delta_{D}^{K\pi\pi^{0}});
\end{eqnarray}
the second is the direct \CP asymmetry,
\begin{eqnarray}
{ A_{DK} } & = & {\frac{\mathcal{B}(
[K^{+}\pi^{-}\pi^{0}]_{D}K^{-})-\mathcal{B}(
[K^{-}\pi^{+}\pi^{0}]_{D}K^{+})}{\mathcal{B}(
[K^{+}\pi^{-}\pi^{0}]_{D}K^{-})+\mathcal{B}(
[K^{-}\pi^{+}\pi^{0}]_{D}K^{+})}}\\ \nonumber
 & = &  \frac{2r_{B}r_{D}R_{K\pi \pi^{0}}\sin \phi_{3}\sin(\delta_{B}+ 
\delta_{D}^{K\pi\pi^{0}})} {r_{B}^{2}+r_{D}^{2}+2r_{B}r_{D}R_{K\pi
\pi^{0}}\cos \phi_{3} \cos (\delta_{B}+ \delta_{D}^{K\pi\pi^{0}})},
\end{eqnarray}

where $r_B$ and $\delta_B$ are the absolute ratio and strong-phase
difference between the suppressed $B^{-}\to \Dbar^0K^{-}$ decay and
the favored  $B^{-}\to D^{0}K^{-}$ decay amplitudes. Furthermore,
the ratio of DCS and CF $D$ decays $r_D$ is defined via
 \begin{equation} r_{D}^{2}  \equiv \frac{\Gamma(D^{0}\rightarrow {K^{+}}\pi^{-}\pi^{0})}{\Gamma(D^{0}\rightarrow
 K^{-}\pi^{+}\pi^{0})} = 
\frac{\int{d\overrightarrow{\mathbf{m}}A^{2}_{DCS}(\overrightarrow{\mathbf{m}})}}{\int{d\overrightarrow{\mathbf{m}}
A^{2}_{CF}(\overrightarrow{\mathbf{m}})}},
 \end{equation}
and the coherence factor $R_{K\pi\pi^{0}}$ and average strong-phase
difference $\delta_{D}^{K\pi\pi^{0}}$ \cite{AS} via
 \begin{equation} R_{K\pi\pi^{0}}e^{i\delta_{D}^{K\pi\pi^{0}}} \equiv 
\frac{\int{d\overrightarrow{\mathbf{m}}A_{DCS}(\overrightarrow{\mathbf{m}})A_{CF}(\overrightarrow{\mathbf{m}})e^{i\delta
(\overrightarrow{\mathbf{m}})}}}
 {\sqrt{\int{d\overrightarrow{\mathbf{m}}A^{2}_{DCS}(\overrightarrow{\mathbf{m}})}\int{d\overrightarrow{\mathbf{m}}A^{2}_{CF}(\overrightarrow{\mathbf{m}})}}}.
 \end{equation}
 Here, $A_{CF}(\overrightarrow{\mathbf{m}})$ and $A_{DCS}(\overrightarrow{\mathbf{m}})$ are the magnitudes of the
CF and DCS amplitudes, respectively,
$\delta(\overrightarrow{\mathbf{m}})$ is the relative strong phase,
and $\overrightarrow{\mathbf{m}} \equiv [m_{K\pi}^2,\,
m_{K\pi^0}^2]$ indicates a point in the Dalitz plane.

The definition of $R_{K\pi\pi^{0}}$ is such that its value is
bounded between zero and one. Sensitivity to $\phi_3$ through
measurements of $R_{DK}$ and $A_{DK}$ is maximal when
$R_{K\pi\pi^{0}}$ is unity. The measured value of $R_{K\pi\pi^{0}}$
is $0.84\pm0.07$ \cite{cleo}, which means that these observables are
suitable to obtain information about $\phi_3$. The previous
measurement of this channel \cite{babar1} has constrained $R_{DK}$
to be less than $2.1\times 10^{-2}$ at the 90\% confidence level; no
limit on $A_{DK}$ is presented.

The observables for the $B^{-}\to D\pi^{-}$ mode are $R_{D\pi}$ and
$A_{D\pi}$. They can be defined using Eqs. (1) and (2) with the
following substitutions: $K\to\pi$ for the $B$ daughter, $r_B\to
r_B^{D\pi}$, and $\delta_B \to \delta_B^{D\pi}$. Here, $r_B^{D\pi}$
and $\delta_{B}^{D\pi}$ are the absolute ratio and strong-phase
difference between the suppressed and favored $B^{-}\to D\pi^{-}$
decay amplitudes. The sensitivity to $\phi_3$ is reduced in this
mode because $r_B^{D\pi}$ is approximately an order of magnitude
smaller than $r_B$. There have been no previous measurements of
$R_{D\pi}$ and $A_{D\pi}$. In Ref. \cite{rama} it has been shown that the corrections due to $D$-mixing on $R_{D\pi}$ and $A_{D\pi}$ 
are potentially large; therefore, such corrections would need to be taken into account
if these measurements are used in the determination of $\phi_3$.

Our measurement uses a data sample of $772 \times 10^6 B\Bbar$
pairs, collected with the Belle detector \cite{detector} located at
the KEKB asymmetric-energy $e^+e^-$ (3.5 on 8~GeV) collider
\cite{collider} operating near the $\Upsilon(4S)$ resonance. The
principal detector elements used in this analysis are a silicon
vertex detector, a 50-layer central drift chamber (CDC), an array of
aerogel threshold Cherenkov counters (ACC),  
a barrel-like arrangement of time-of-flight scintillation counters
(TOF), and an electromagnetic calorimeter comprised of CsI(Tl)
crystals located inside a super-conducting solenoid coil that
provides a 1.5~T magnetic field.

 We reconstruct $\pi^{0}$ candidates
from photon pairs that have a momentum greater than 400~MeV/$c$ in
the $e^+e^-$ center-of-mass (CM) frame and an invariant mass between
120 and 145~MeV/$c^2$, which corresponds to approximately
$\pm3.2\sigma$ in resolution around the nominal $\pi^0$ mass
\cite{pdg}. Each photon candidate is required to have an energy
greater than 50~MeV. We apply a mass-constrained fit to the
$\pi^{0}$ candidate to improve its momentum resolution.

Neutral $D$ meson candidates are reconstructed from a pair of
oppositely charged tracks and a $\pi^{0}$ candidate. Each track must
have a distance of closest approach to the interaction point of less
than 0.2~cm in the plane transverse to the positron beam direction
and less than 1.5~cm along the positron beam axis. We also define
$L_{K}$ ($L_{\pi}$), the likelihood of a track being a kaon (pion),
based on particle identification (PID) information \cite{pid} from
the ACC and the TOF, combined with specific ionization measured in
the CDC. We apply likelihood-ratio requirements of $L(K/\pi) =
\frac{L_{K}}{L_{K}+L_{\pi}} > 0.6$ for a kaon candidate and
$L(K/\pi) < 0.4$ for a pion candidate. The efficiency to identify a
kaon (pion) is approximately 83\% (88\%) averaged over momentum and
the probability of misidentifying a pion (kaon) as a kaon (pion) is
approximately 8\% (7\%). The invariant mass of $K\pi\pi^0$
candidates is required to satisfy $1.804 ~\mathrm{GeV}/c^2 <
M_{K\pi\pi^0} < 1.885 ~\mathrm{GeV}/c^2$, which corresponds to
approximately $\pm2.5\sigma$ in resolution around the nominal $D$
mass \cite{pdg}. To improve the four momentum resolution of the
daughters, we apply a $D$-mass constrained fit.

A $B$ meson candidate is reconstructed by combining the $D$
candidate with a charged hadron. The same set of $L(K/\pi)$
requirements is applied for the prompt track as that used for $D$
meson reconstruction. The signal is identified with the
beam-energy-constrained mass $M_{\rm bc} = c^{-2}\sqrt{E^{2}_{\rm beam} - |\vec{p}_{B}|^{2}c^{2}}$ and the energy 
difference $\Delta E = E_{B}- E_{\rm beam}$, where $E_{\rm beam}$ is
the beam energy and $\vec{p}_{B}$ $(E_{B})$ is the momentum (energy)
of the $B$ meson candidates in the CM frame. For $B\to DK$ decays,
$M_{\rm bc}$ peaks at the nominal mass of the $B$ meson \cite{pdg}
and $\Delta E$ peaks at zero. We select candidates in the ranges $
5.27~\mathrm{GeV}/c^{2} < M_{\rm bc} < 5.29~\mathrm{GeV}/c^{2}$ and
$ -0.1~\mathrm{GeV} < \Delta E < 0.2~\mathrm{GeV} $.

To suppress background coming from the ${D^{*}}^{\pm}\rightarrow
D\pi^{\pm}$ decays in $e^{+}e^{-}\rightarrow c\bar{c}$, we use the
mass difference between the $D^{*\pm}$ and $D$ candidates
$\left(\Delta M\right)$. We reconstruct $D^{*\pm}$ candidates from
the $D$ meson used for $B$ reconstruction and a $\pi^{\pm}$
candidate not used in the $B$ reconstruction. No PID requirement is
applied to the $\pi^{\pm}$ because of its low momentum when coming
from the ${D^{*}}^{\pm}$ decay. After requiring $\Delta M >
0.15~\mathrm{GeV}/c^{2}$, we remove $99\%$ of ${D^{*}}^{\pm}$
backgrounds and $17\%$ of all $c\bar{c}$ backgrounds. The relative
loss of signal efficiency is $3.4\%$.

A possible source of peaking background is the favored
$B^-\to[K^-\pi^+\pi^0]_Dh^-$ ($h = K$ or $\pi$) decay, which can
contribute to the signal region of the respective suppressed decay,
due to misidentification of both the $K^{-}$ and $\pi^{+}$ mesons in
the $D$ decay. To reject this background, we veto events satisfying
$1.804 ~\mathrm{GeV}/c^2 < M_{K\pi\pi^0} < 1.885 ~\mathrm{GeV}/c^2$
when the mass assignments of the $K^{-}$ and $\pi^{+}$ are
exchanged. This criterion reduces the background to a negligible
level with a relative loss of signal efficiency of around $17\%$.
About 6\% of events have multiple $B$ candidates; the candidate with
$M_{K\pi\pi^0}$ and $M_{\rm bc}$ most consistent with the
corresponding nominal values is retained for further analysis.

The dominant remaining background for both the favored and the
suppressed $Dh$ decays comes from $e^{+}e^{-}\rightarrow q\bar{q}$
($q$ = $u$, $d$, $s$, or $c$) continuum events. The daughters from
$B\Bbar$ events tend to emerge isotropically in the CM frame whereas
the particles from continuum events are collimated into back-to-back
jets. We exploit this difference in event topology by using a neural
network \cite{BelleADS,NN} to combine shape variables that describe
the particle distribution with other properties of the event that
differentiate between $q\bar{q}$ and $B\Bbar$ events.

The neural network utilizes the following nine input variables: 1)
the likelihood ratio of the Fisher discriminant formed from 17
modified Fox-Wolfram moments \cite{sfw}; 2) the absolute value of
the cosine of the angle in the CM frame between the thrust axis of
the $B$ decay and that of the remaining particles in the event; 3)
the vertex separation between the $B$ candidate and the remaining
charged tracks along the beam direction; 4) the cosine of the angle
between the direction of the $K$ candidate from the $D$ decay and
the direction opposite the flight of the $B$ candidate measured in
the $D$ rest frame; 5) the absolute value of the $B$ flavor tagging
dilution factor \cite{ftag}; 6) the cosine of the angle between the
$B$ flight direction and the beam axis in the CM frame; 7) the
cosine of the angle between the $D$ and $\Upsilon(4S)$ directions in
the rest frame of the $B$; 8) the product of the charge of the $B$
candidate and the sum of the charges of all kaons not used for the
reconstruction of the $B$ candidate; and 9) the difference between
the sum of the charges of particles in the $D$ hemisphere and the
sum of charges in the opposite hemisphere, excluding the particles
used in the $B$ meson reconstruction.

The neural network output $\cal{C_{\mathrm{NB}}}$ is in the range
$-1$ to 1, where events at $\cal{C_{\mathrm{NB}}}$ = 1 $(-1)$ are
signal (continuum) like. The training and optimization of the neural
network are carried out with signal and $q\bar{q}$ Monte Carlo (MC)
samples after event-selection requirements are imposed. We require
$\cal{C_{\mathrm{NB}}}$ $>$ $-0.6$, which rejects $70\%$ of the
$q\bar{q}$ continuum background and only $3\%$ of the signal. The
selection efficiency after all criteria have been applied is
$10.9\%$ ($11.2\%$) for $B\to DK$ ($B\to D\pi$) decays.

The $\cal{C_{\mathrm{NB}}}$ distribution peaks strongly at
$|\cal{C_{\mathrm{NB}}}|$ $\sim$ 1 and is therefore difficult to
model with a simple analytic function. Therefore, to improve this
modeling, we transform $\cal{C_{\mathrm{NB}}}$ to a new variable
$\cal{C'_{\mathrm{NB}}}$:
\begin{eqnarray}
 \cal{C'_{\mathrm{NB}}} &=& \log\left(\frac{\cal{C_{\mathrm{NB}}} - \cal{C_{\mathrm{NB,min}}}}{\cal{C_{\mathrm{NB,max}}} - \cal{C_{\mathrm{NB}}}}\right) \; .
\end{eqnarray}
Here, $\cal{C_{\mathrm{NB,min}}}$ = $-0.6$ and $\cal{C_{\mathrm{NB,max}}}$ = 1 
are the minimum and maximum values of $\cal{C_{\mathrm{NB}}}$ for
the events used for the signal extraction. The distribution of
$\cal{C'_{\mathrm{NB}}}$ can be modeled by Gaussian or asymmetric
Gaussian functions.

We extract the signal yield using an unbinned extended maximum
likelihood fit to $\Delta E$ and $\cal{C'_{\mathrm{NB}}}$
distributions. We perform separate fits to the suppressed and
favored $B\to DK$ ($B\to D\pi$) modes. The total PDF for each
component is formed by multiplying the individual PDFs for $\Delta
E$ and $\cal{C'_{\mathrm{NB}}}$, as they have negligible
correlation. The $\Delta E$ and $\cal{C'_{\mathrm{NB}}}$ PDF for
each fit component are described as follows.

For signal, the $\Delta E$ distribution is parameterized by a sum of
two Gaussian functions of common mean. The $\cal{C'_{\mathrm{NB}}}$
distribution is parameterized by the sum of a symmetric Gaussian and
an asymmetric Gaussian having different means. The PDF shape
parameters used in the fit to the suppressed mode are fixed to the
values obtained from the fit to the favored mode.

For $B\to DK$ decays, there is a background from $B\to D\pi$ decays
where the $\pi$ daughter of the $B$ is misidentified as a $K$. This
background peaks in $\Delta E$ at around 45~MeV and is modeled by
the sum of a symmetric Gaussian and an asymmetric Gaussian. The
distribution of $\cal{C'_{\mathrm{NB}}}$ is the same as for the
signal, so the same PDF is used. For the fit to the suppressed $DK$
data, the $D\pi$ background yield is fixed to that measured in the
suppressed $D\pi$ signal fit multiplied by the misidentification
rate; this procedure reduces the statistical uncertainty on the
signal yield. For this background component, all other PDF shape
parameters in the suppressed mode are fixed to those measured in the
fit to the favored mode.

The $B\Bbar$ background in the favored $Dh$ modes has two
components. The first is from $B^-\rightarrow D^{*}h^-$ and
$B^-\rightarrow D\rho^-$ events and peaks at $\Delta E<
-0.1~\mathrm{GeV}$, so an upper tail is observed within the fit
range. The second component is combinatorial. The peaking and
combinatorial components are modeled by an exponential and
first-order polynomial, respectively. The suppressed $Dh$ has a much
smaller peaking $B\Bbar$ background contribution than the favored
mode, so an exponential function is used to model the whole peaking
and combinatorial background. The $\cal{C'_{\mathrm{NB}}}$
distribution for the $B\Bbar$ background is parameterized by a
Gaussian function, which is determined separately for suppressed and
favored modes from the $B\Bbar$ MC sample.

The $\Delta E$ and $\cal{C'_{\mathrm{NB}}}$ distributions for the
$q\bar{q}$ continuum background are parametrized by a first-order
polynomial and a sum of two Gaussian functions of common mean,
respectively. The parameters for $\cal{C'_{\mathrm{NB}}}$ are
determined using the $M_{\rm bc}$ sideband, given by $5.20
~\mathrm{GeV}/c^2 < M_{\rm{bc}} < 5.24 ~\mathrm{GeV}/c^2$, for all
modes.  For the suppressed mode, the mean of one of the Gaussians is
left free in the fit to data; this minimizes the cross feed between
the $q\bar{q}$ and combinatorial $B\Bbar$ backgrounds.

The projections of the fits for the suppressed and favored $Dh$
modes are shown in Figs.~\ref{fig:fit} and~\ref{fig:fit1},
respectively. Suppressed $DK$ and $D\pi$ signal peaks are visible.
The values of $R_{Dh}$ are determined using the signal yields and
efficiencies given in Table~\ref{tab:result}:

\begin{figure}[htbp]
\begin{center}
\begin{tabular}{cccc}
  \label{fig:asym1}\includegraphics[width=0.25\textwidth]{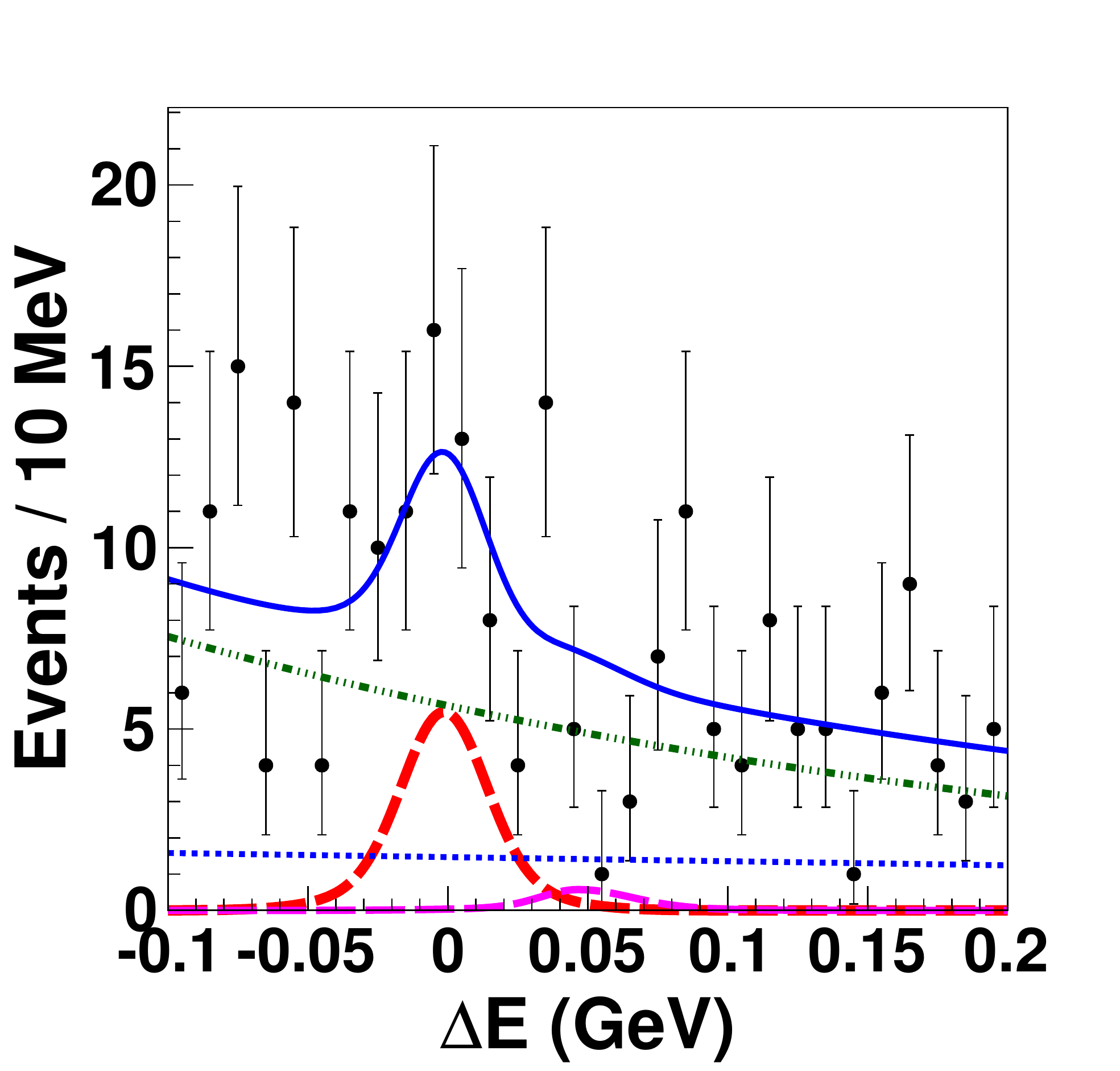}&
  \label{fig:asym2}\includegraphics[width=0.25\textwidth]{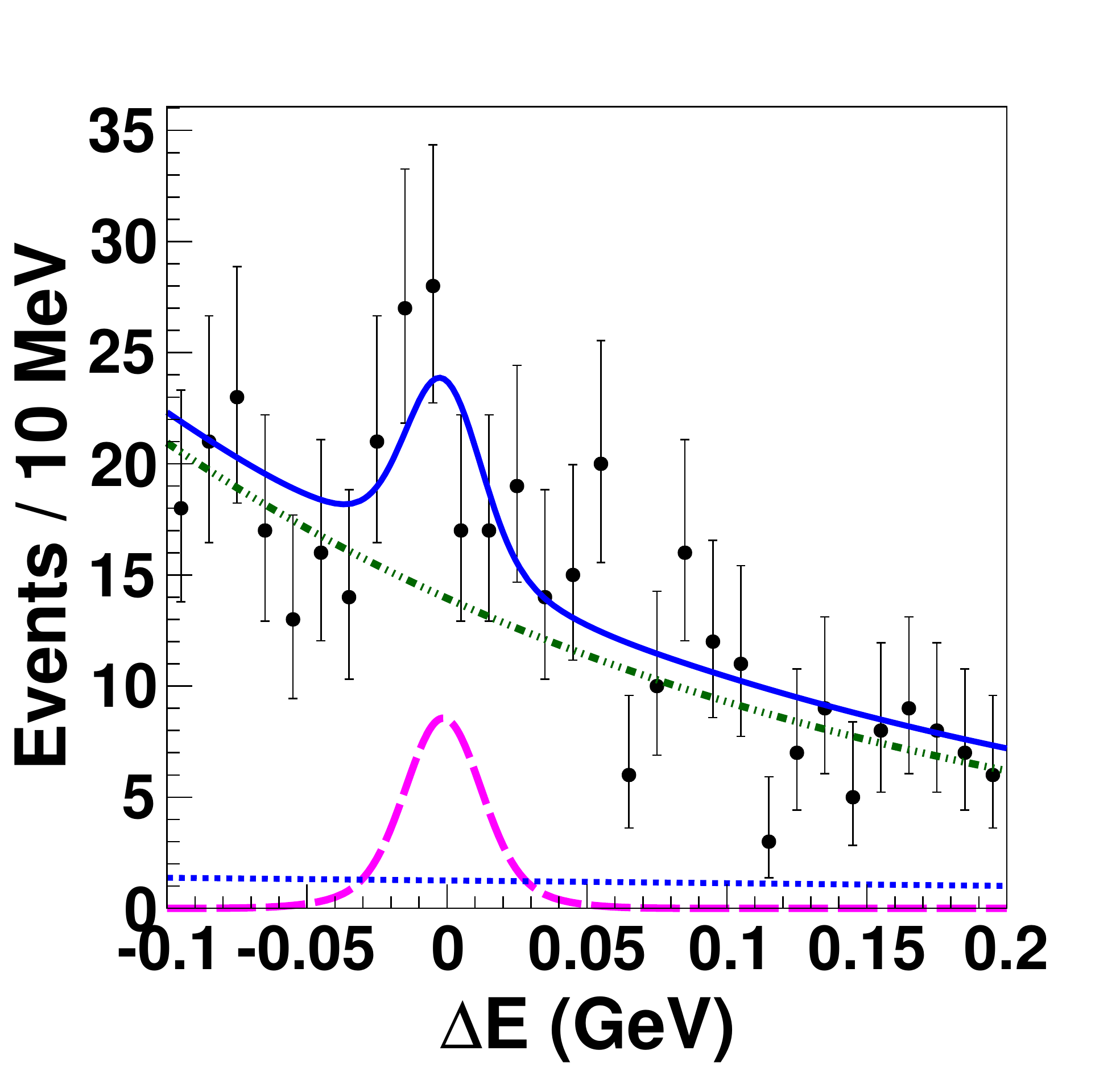}\\
  \label{fig:asym3}\includegraphics[width=0.25\textwidth]{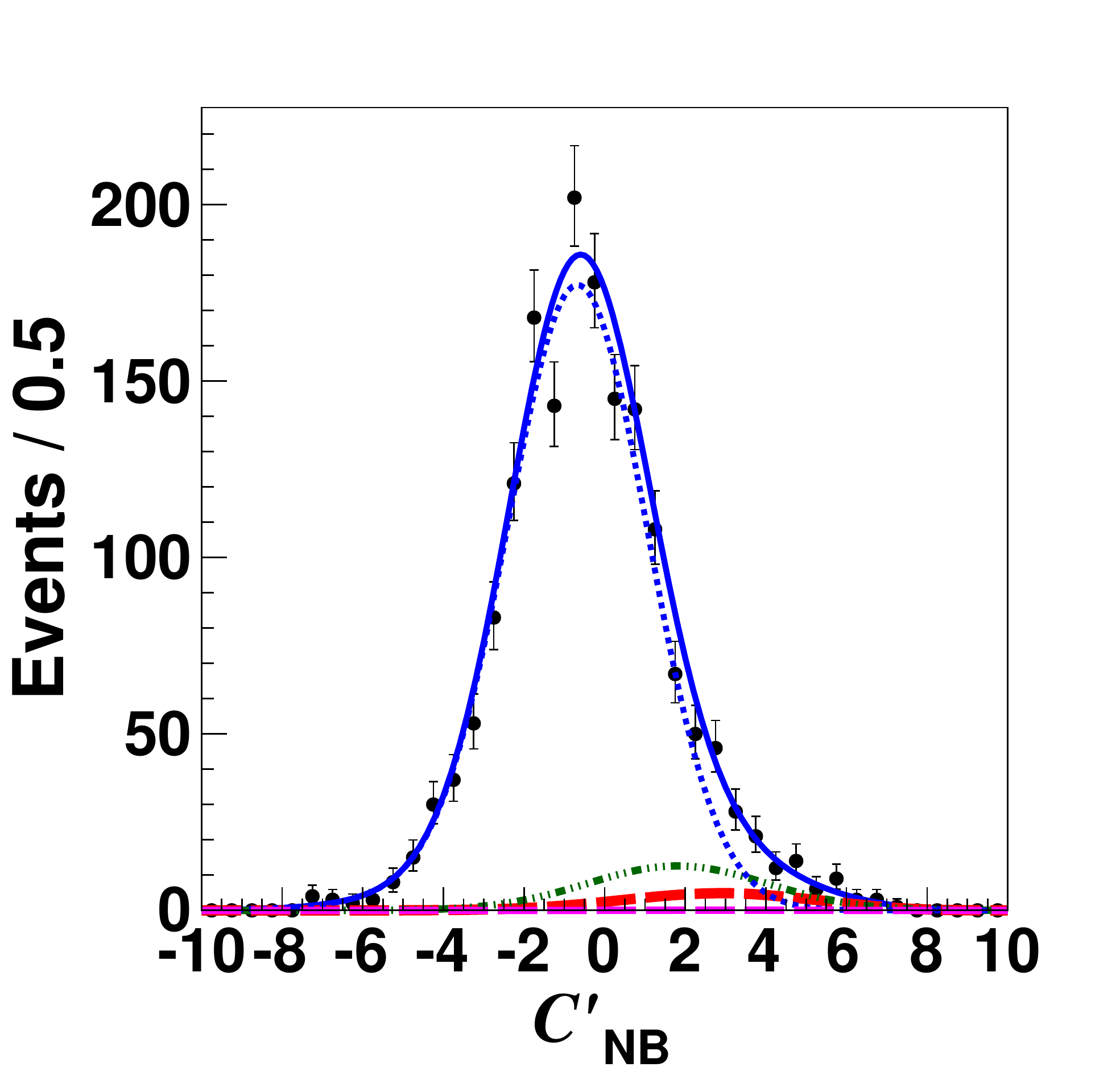}&
  \label{fig:asym4}\includegraphics[width=0.25\textwidth]{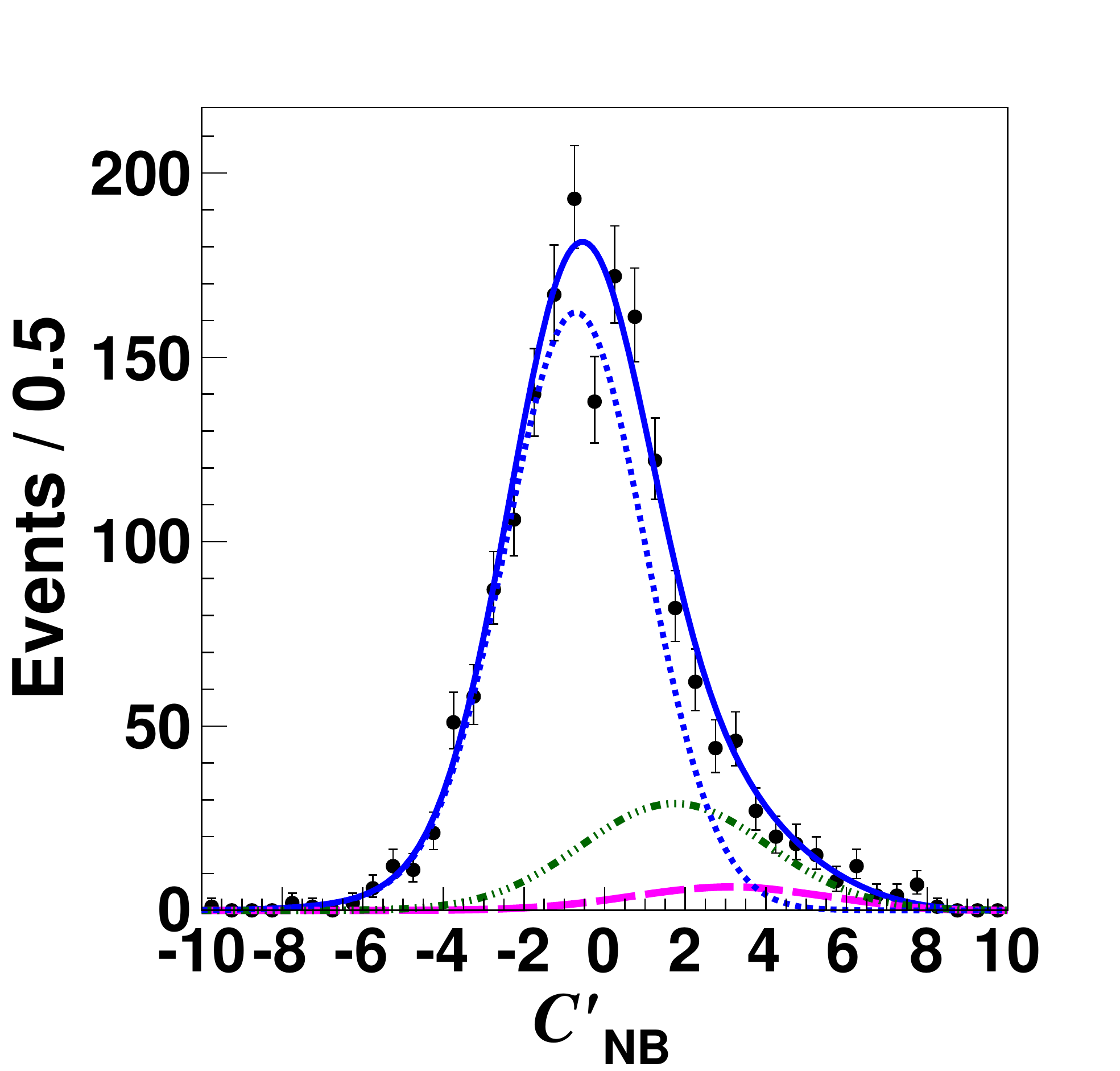}\\
   \end{tabular}
\caption{[color online].~$\Delta E$~($\cal{C'_{\mathrm{NB}}}$ $>$
$4$) and $\cal{C'_{\mathrm{NB}}}$ ($|\Delta E| < 0.02~\mathrm{GeV}$)
distributions for $[K^+\pi^-\pi^0]_{D}K^-$ (left),
$[K^+\pi^-\pi^0]_{D}\pi^-$ (right). In these plots, points with
error bars represent data while the total best-fit projection is
shown with the solid blue curve, for which the components are shown
with thicker dashed red ($DK$ signal), thinner dashed magenta
($D\pi$), dashed dot green ($B\Bbar$ background) and dotted blue
($q\bar{q}$ background). To enhance the signal and suppress the
dominant continuum background in the $\Delta E$ projection, a strict
criterion on $\cal{C'_{\mathrm{NB}}}$ is applied.}
 \label{fig:fit}
 \end{center}
\end{figure}
\begin{figure}[htbp]
\begin{center}
\begin{tabular}{cccc}
  \label{fig:asym1}\includegraphics[width=0.25\textwidth]{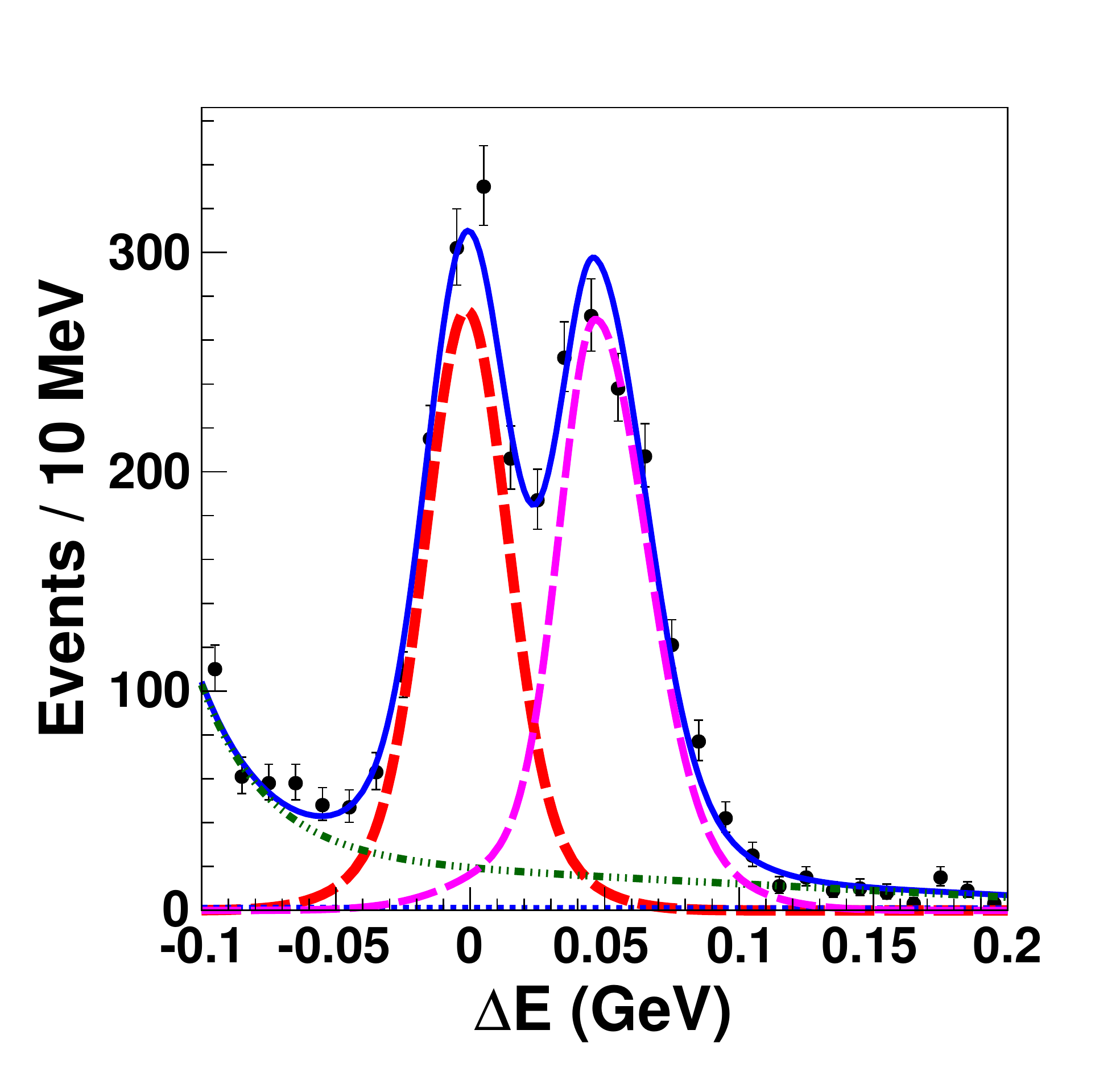}&
  \label{fig:asym2}\includegraphics[width=0.25\textwidth]{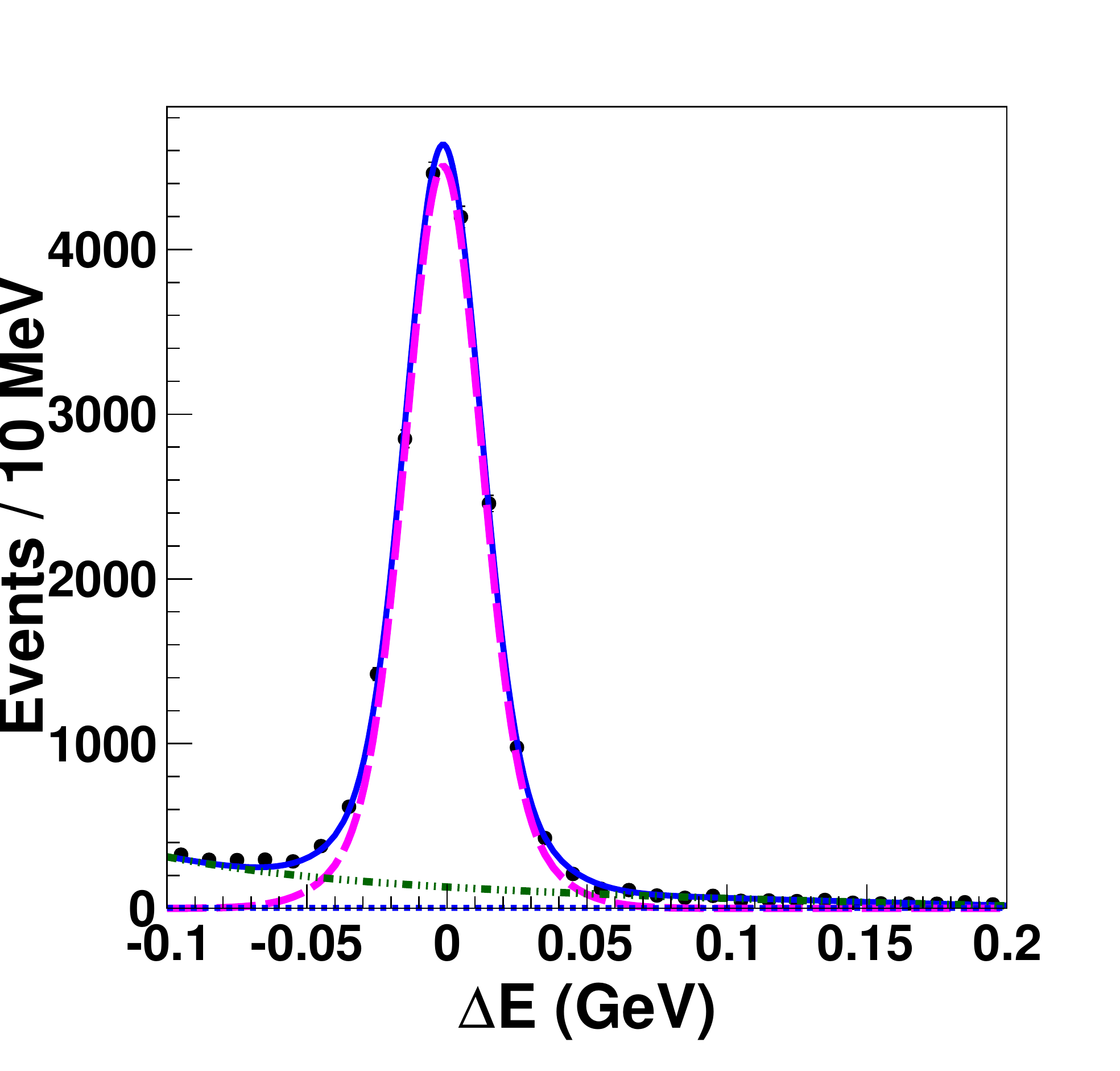}\\
  \label{fig:asym3}\includegraphics[width=0.25\textwidth]{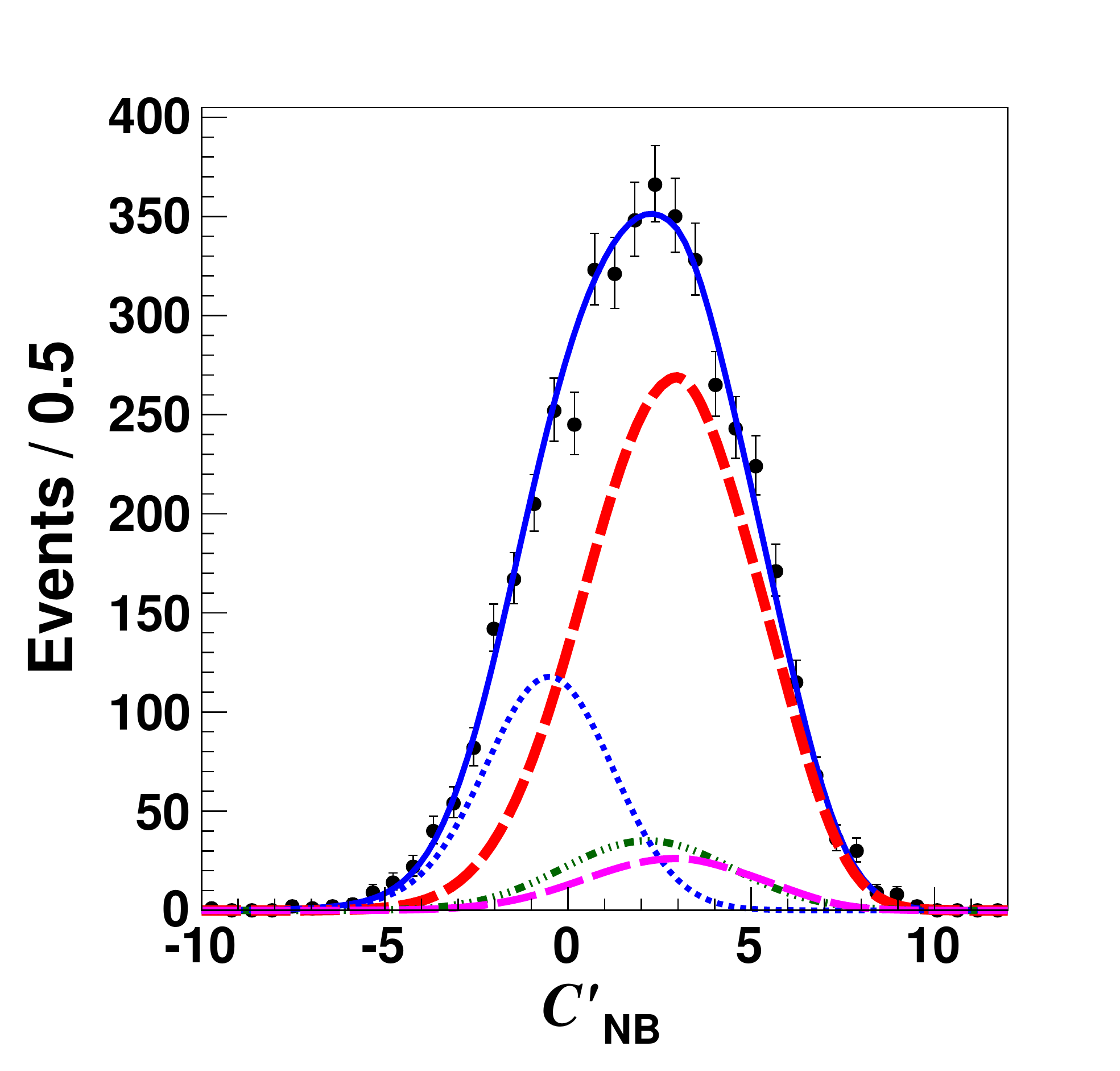}&
  \label{fig:asym4}\includegraphics[width=0.25\textwidth]{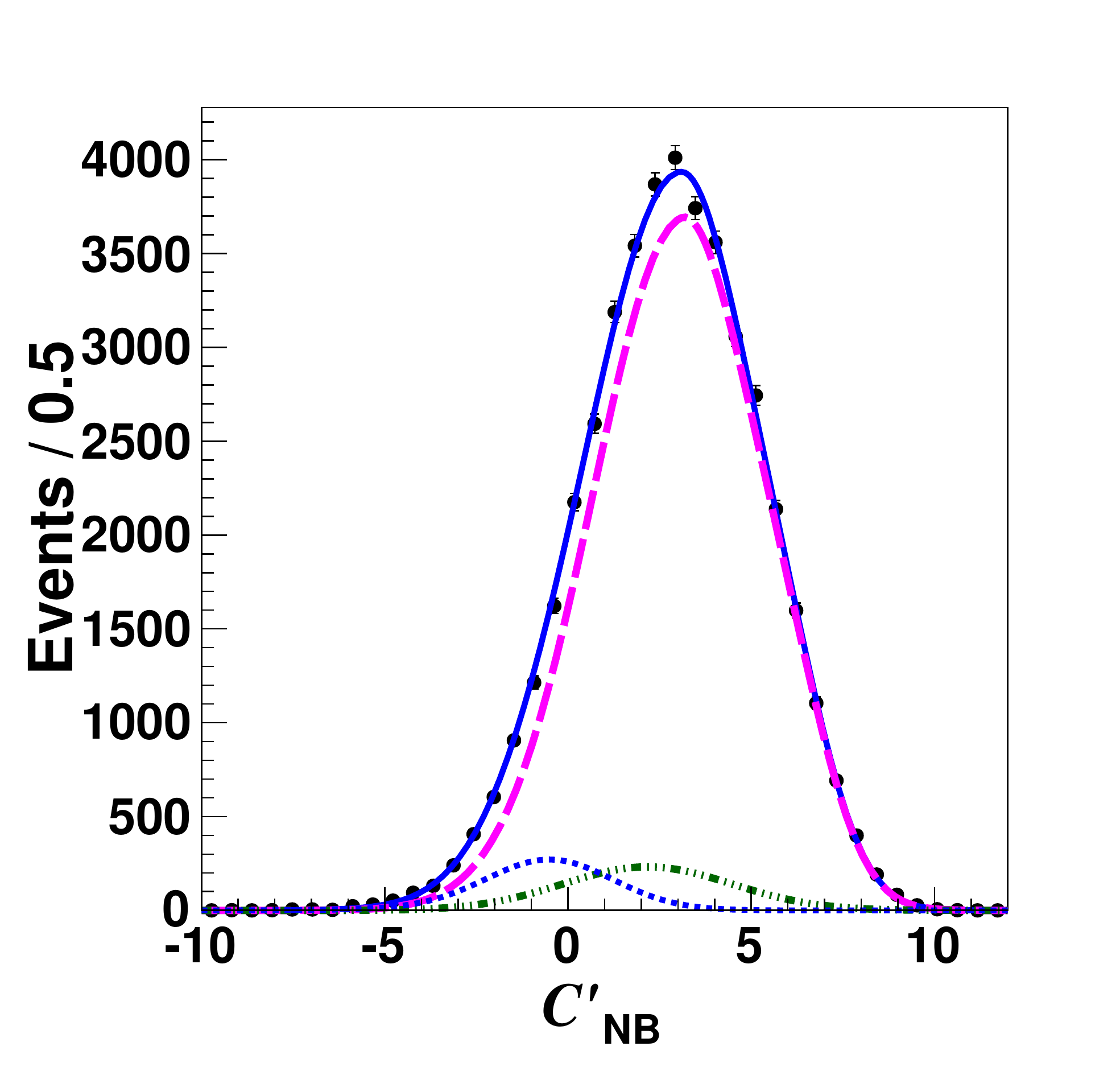}\\
   \end{tabular}
\caption{[color online].~$\Delta E$~($\cal{C'_{\mathrm{NB}}}$ $>$
$4$) and $\cal{C'_{\mathrm{NB}}}$ ($|\Delta E| < 0.02~\mathrm{GeV}$)
distributions for $[K^-\pi^+\pi^0]_{D}K^-$ (left),
$[K^-\pi^+\pi^0]_{D}\pi^-$ (right). The color legend and fit
components are the same as those in Fig.~\ref{fig:fit}.}
 \label{fig:fit1}
 \end{center}
\end{figure}
\begin{eqnarray}
R_{DK} & = &[1.98\pm0.62
(\mathrm{stat.})\pm0.24(\mathrm{syst.})]\times10^{-2}, \\
 R_{D\pi}& = &[1.89\pm0.54
(\mathrm{stat.})^{+0.22}_{-0.25}(\mathrm{syst.})]\times10^{-3}.
\end{eqnarray}

The systematic uncertainties associated with $R_{DK}$ and $R_{D\pi}$
are are listed in Table \ref{tab:syst} and estimated as follows. The
uncertainties due to fixed PDF shape parameters that are obtained
from data are estimated by varying each fixed parameter by
$\pm1\sigma$. The uncertainty due to the $B\Bbar$
$\cal{C'_{\mathrm{NB}}}$ PDF is estimated by varying the mean and
width of the Gaussian by the maximum differences observed between
data and MC for the $\cal{C'_{\mathrm{NB}}}$ PDF from favored
signal. Possible bias related to the fit is checked with 10000
simulated experiments. No bias is observed, and the systematic
uncertainty due to possible bias is taken to be the error on the
mean residual. A small bias is observed in the yields of $B\Bbar$
and $q\bar{q}$ backgrounds in the suppressed $B\to DK$ mode
simulations. This is due to an imperfect modeling of the continuum
$\cal{C'_{\mathrm{NB}}}$ distribution in the signal region by the
fits to the $M_{\rm bc}$ sideband. The impact of this bias on the
signal yield is estimated using simulated experiments to be at most
$3\%$.

Charmless $B^{-}\to K^{-}K^{+}\pi^{-}\pi^{0}$ decay could result in
an irreducible peaking background to the signal. The size of this
background is bounded by fits to the sidebands of the reconstructed
$D$ mass:~$1.45~\mathrm{GeV}/c^2 < M_D < 1.80 ~\mathrm{GeV}/c^2$ and
$1.90~\mathrm{GeV}/c^2 < M_D < 2.25 ~\mathrm{GeV}/c^2$. We apply the
same fitting method used in the signal extraction to the sideband
sample to obtain an expected yield of $-9\pm7$ and $-11\pm8$ events
for suppressed $DK$ and $D\pi$, respectively. Since the yields are
consistent with zero, we include the uncertainty on the obtained
yield as a systematic uncertainty. This is the dominant source of
systematic uncertainty on the measurement of $R_{DK}$.

There are also uncertainties on the efficiency coming from the
limited statistics of the MC sample and the calibration of the PID
efficiency for potential data-MC differences. The uncertainty due to
fixing the $B\to D\pi$ yield in the fit to the suppressed $B\to DK$
sample is found to be negligible.

The signal significance is calculated as
$\mathcal{S}=\sqrt{-2\mathrm{ln(L_{0}/L_{max})}}$, where
$\mathrm{L_{max}}$ is the maximum likelihood and $\mathrm{L_{0}}$ is
the likelihood when the signal yield is constrained to be zero. In
order to include systematic uncertainty in the significance, we
convolve the fit likelihood with a Gaussian whose width is equal to
the systematic uncertainty for $R_{DK}$ and with an asymmetric
Gaussian whose widths are the negative and positive systematic
uncertainties for $R_{D\pi}$. The significance of
$R_{DK}~(R_{D\pi})$ is $3.2\sigma~(3.3\sigma)$.

\begin{table}[htbp]
\begin{center}
\caption{Signal yields, reconstruction efficiencies for signals
after PID calibration for any data-MC discrepancy and significances
($\mathcal{S}$) including systematic uncertainties. The
uncertainties listed for the signal yield are statistical only, and
those on efficiency are from MC statistics and the PID
correction.}\label{tab:result}

\begin{tabular}{lccc}\hline\hline
~~~~~~~Mode & Yield& Efficiency (\%)& $\mathcal{S}$\\
\hline
$B^-\to [K^+\pi^-\pi^0]_{D}K^-$      & 77$\pm$24&10.9$\pm$0.1 &3.2$\sigma$\\
 $B^-\to [K^-\pi^+\pi^0]_{D}K^-$     &3844$\pm$125&10.8$\pm$0.1& \\
$B^-\to [K^+\pi^-\pi^0]_{D}\pi^-$   &94$\pm$27&11.2$\pm$0.1& 3.3$\sigma$  \\
$B^-\to [K^-\pi^+\pi^0]_{D}\pi^-$    & 49668$\pm$338& 11.2$\pm$0.1&  \\

\hline\hline
\end{tabular}
\end{center}
\end{table}

\begin{figure}[htbp]
\begin{center}
\begin{tabular}{cccc}
  \label{fig:asym1}\includegraphics[width=0.25\textwidth]{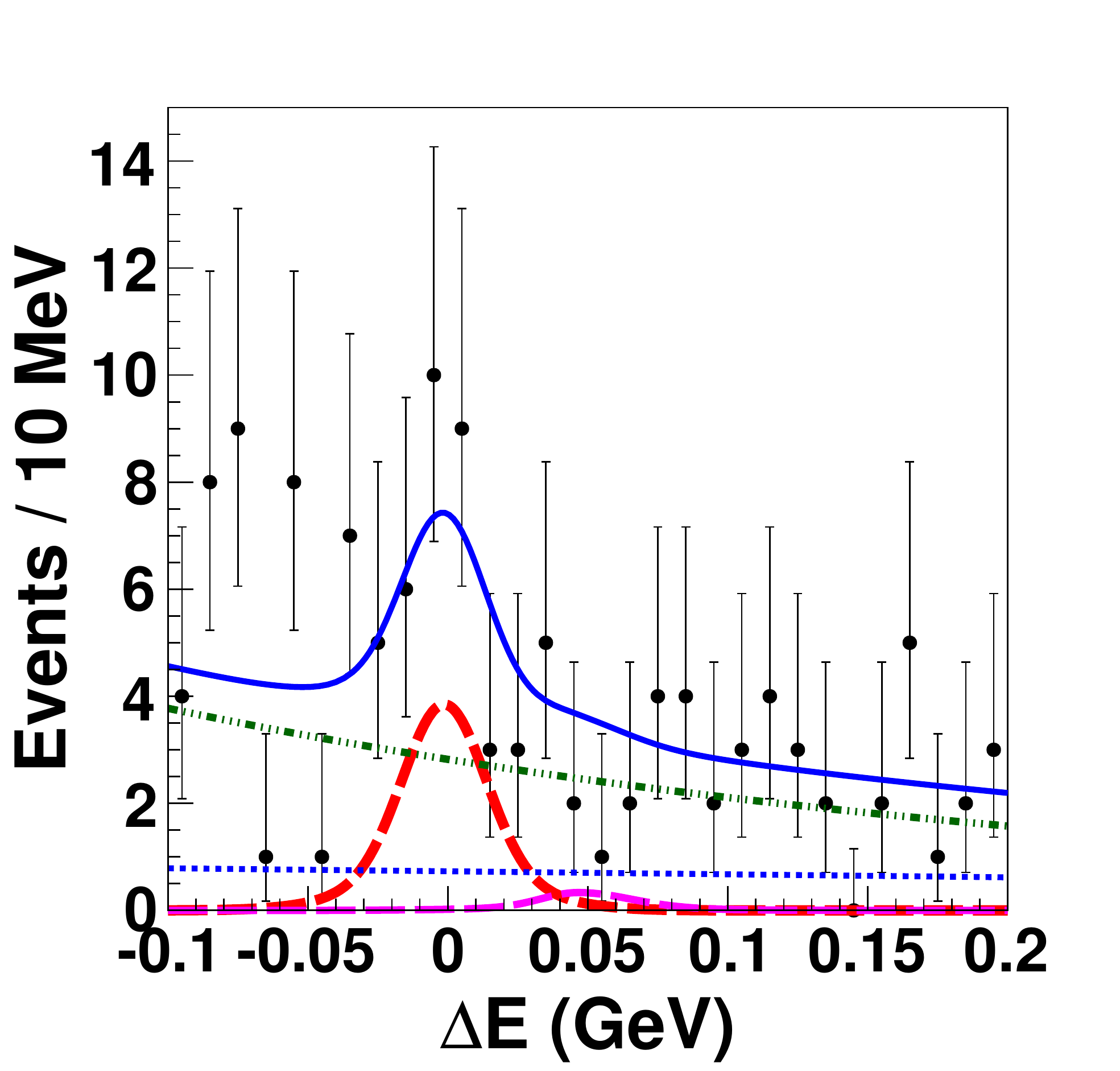}&
  \label{fig:asym2}\includegraphics[width=0.25\textwidth]{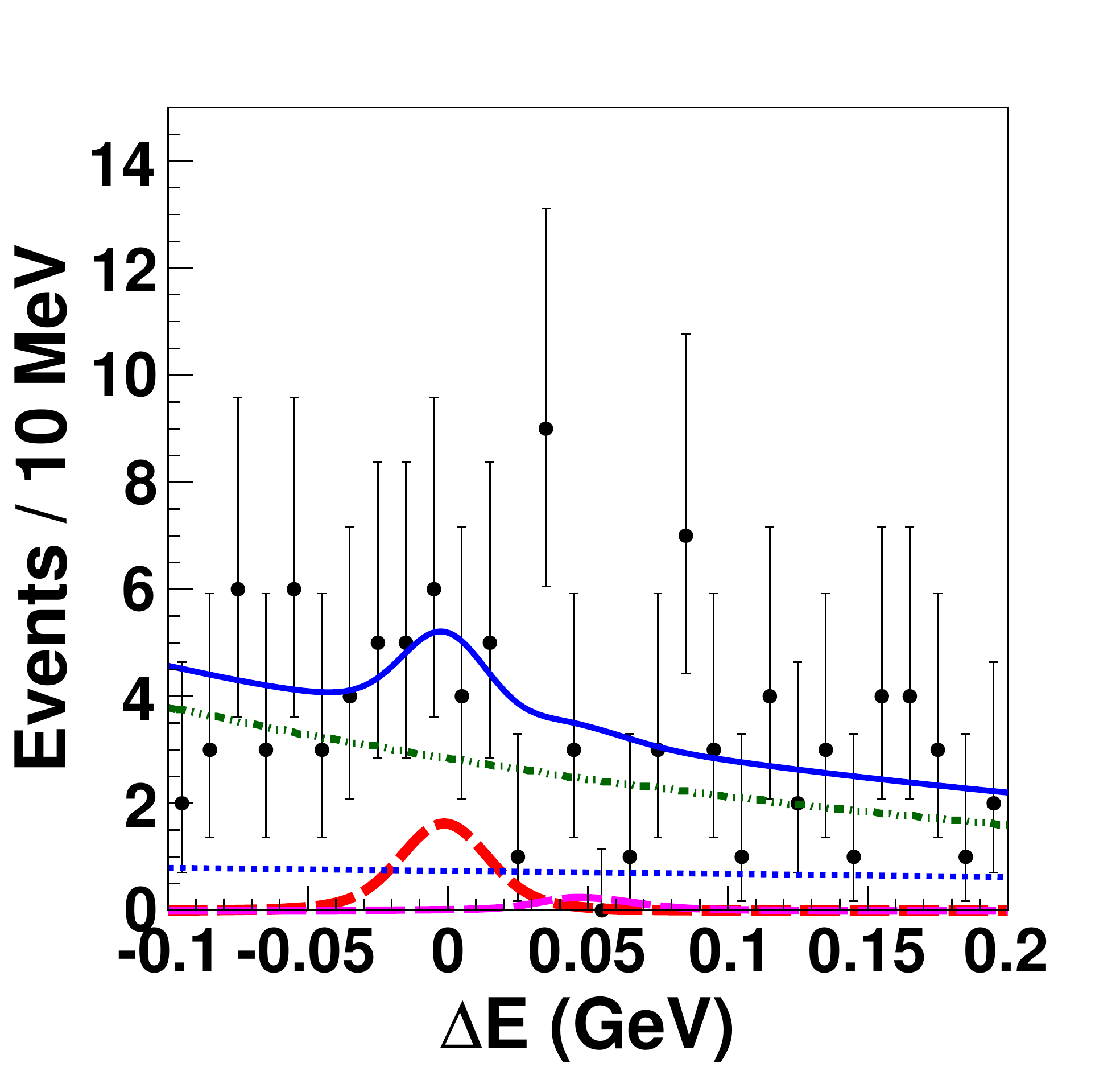}\\
  \label{fig:asym3}\includegraphics[width=0.25\textwidth]{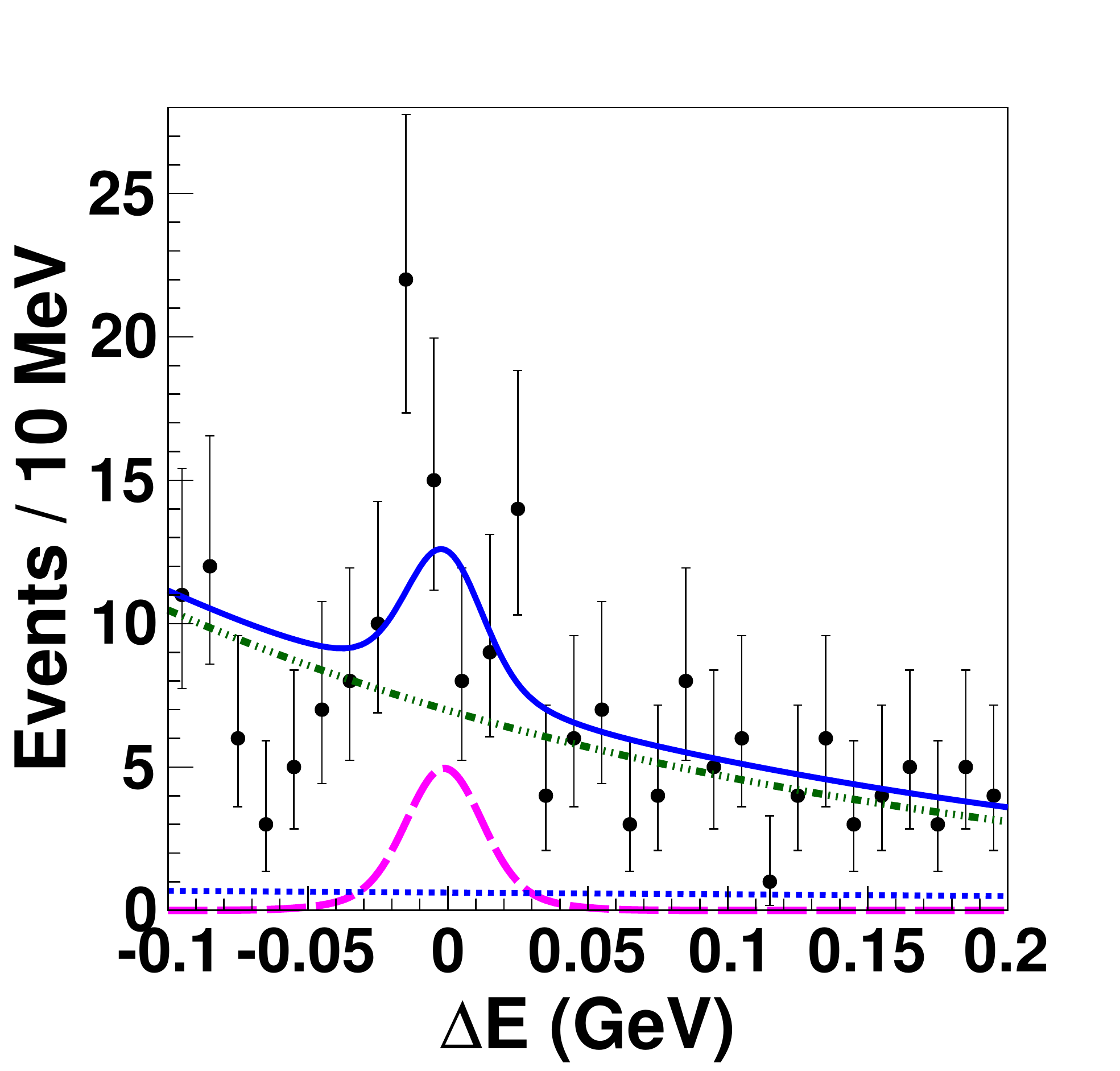}&
  \label{fig:asym4}\includegraphics[width=0.25\textwidth]{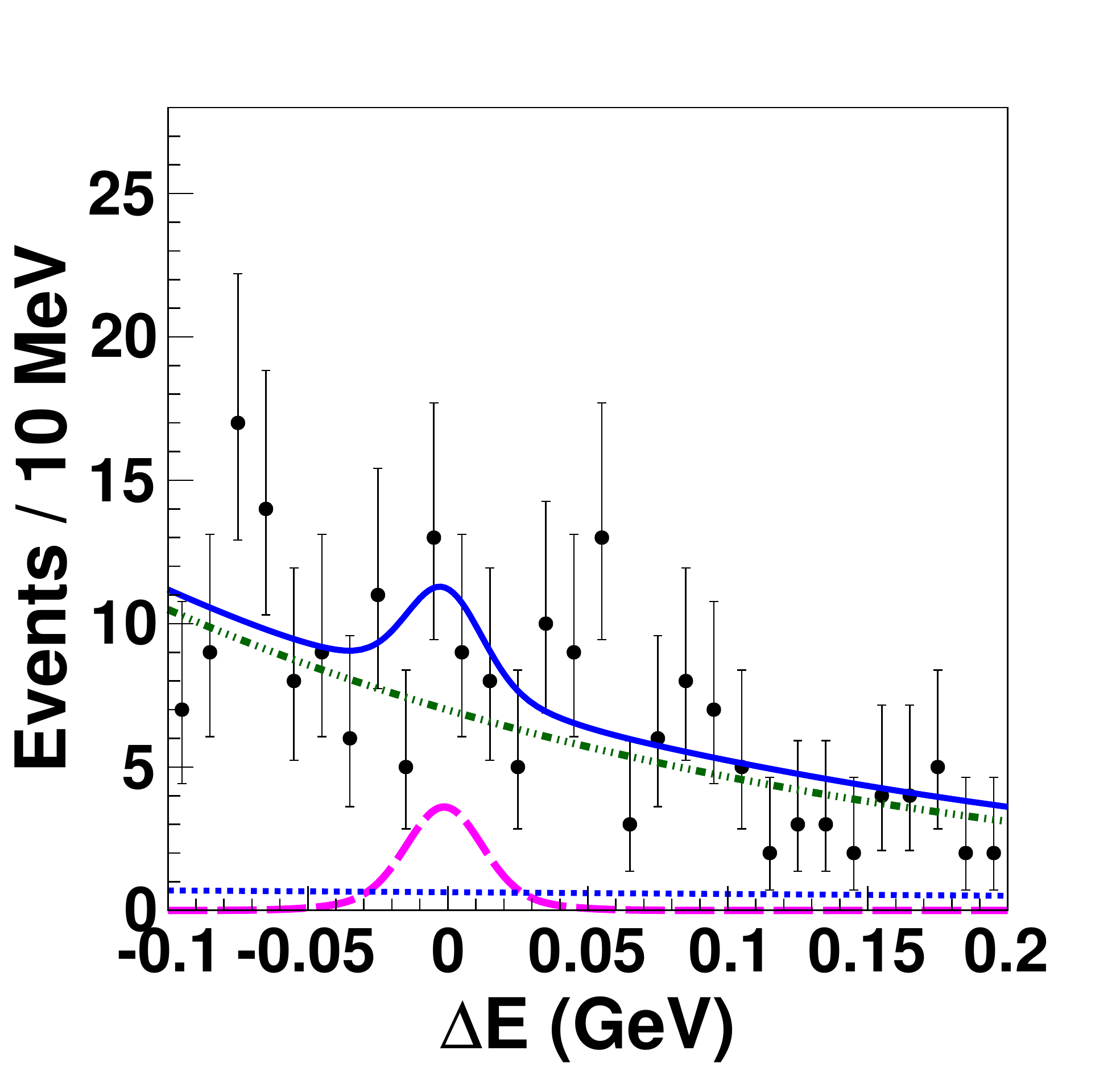}\\
   \end{tabular}
\caption{[color online].~$\Delta E$ distributions
($\cal{C'_{\mathrm{NB}}}$ $>$ $4$) for $[K^+\pi^-\pi^0]_{D}K^-$
(left upper), $[K^-\pi^+\pi^0]_{D}K^+$ (right upper),
$[K^+\pi^-\pi^0]_{D}\pi^-$ (left lower), $[K^-\pi^+\pi^0]_{D}\pi^+$
(right lower). The color legend and fit components are the same as
those in Fig.~\ref{fig:fit}.}
 \label{fig:cp}
 \end{center}
\end{figure}

We measure $A_{Dh}$ in a separate fit to the suppressed candidates,
including the charge of the kaon or pion from the $B$ decay as an
additional observable and $A_{Dh}$ as a new free parameter. Since
asymmetries associated with $B\Bbar$ and $q\bar{q}$ parameters are
expected to be negligible, they are fixed to zero in the $A_{Dh}$
fit. The measured values are:
\begin{eqnarray}
A_{DK} &=& 0.41\pm0.30 (\mathrm{stat.})\pm0.05
(\mathrm{syst.}),\\
A_{D\pi} &=& 0.16\pm0.27 (\mathrm{stat.})^{+0.03}_{-0.04}
(\mathrm{syst.}).
 \end{eqnarray}
The $\Delta E$ projections for signal $Dh^-$ and $Dh^+$ are shown in
Fig.~\ref{fig:cp}. The systematic uncertainties (see Table
\ref{tab:syst}) arise from the following sources. Uncertainties
related to the fit parameters are obtained in the same way as those
estimated for $R_{Dh}$. The uncertainty due to the yield of the
peaking background is $\pm0.04$ ($\pm0.01$) for $A_{DK}$
($A_{D\pi}$), which is estimated under the assumption of zero
asymmetry in the peaking background. A possible bias in $A_{Dh}$ due
to any detector asymmetry is estimated by determining the asymmetry
between $B^{+}$ and $B^{-}$ in the favored mode, which is expected
to be close to zero. No detector asymmetry is observed in the
favored $DK$ mode, so the uncertainty on the measurement is taken as
a systematic uncertainty for the suppressed $DK$ mode. An asymmetry
is seen in the favored $D\pi$ mode, which is taken as a systematic
uncertainty for the suppressed $D\pi$ mode. The remaining sources
are found to give negligible contributions.

\begin{table}[htbp]
\begin{center}
\caption{Summary of the systematic uncertainties for $R_{Dh}$ and
$A_{Dh}$. Negligible contributions are denoted by $``$--$"$.
}\label{tab:syst}

\begin{tabular}{lcccc}\hline\hline
Source &$R_{DK} (\%)$ &$R_{D\pi} (\%)$&$A_{DK}$&$A_{D\pi}$\\
\hline
$\Delta$E and $\cal{C'_{\mathrm{NB}}}$ PDFs &$^{+6.5}_{-7.1}$&$^{+\phantom{0}{8.3}}_{-10.3}$&$^{+0.03}_{-0.02}$&$^{+0.02}_{-0.03}$\\
Fit bias &$+0.1$&$+0.4$ &--&-- \\
Due to $B\Bbar$ and $q\bar{q}$ bias&$\pm3.0$&--&--&--\\
Peaking background &$\pm9.5$&$\pm8.2$&$\pm0.04$&$\pm0.01$\\
Efficiency &$\pm0.1$&$\pm0.1$&--&--  \\
Detector asymmetry &--&--&$\pm0.02$&$\pm0.02$  \\
\hline
Total&$^{+11.9}_{-12.2}$&$^{+11.7}_{-13.2}$&$\pm0.05$&$^{+0.03}_{-0.04}$\\
 \hline\hline
\end{tabular}
\end{center}
\end{table}

In summary, for the mode $B^-\to Dh^{-}$, $D\to K^{+}\pi^{-}\pi^{0}$
($h=K,\pi$), we report the measurements $R_{Dh}$ and $A_{Dh}$, using
$772 \times 10^6$ $B\Bbar$ pairs collected
by the Belle detector. 
We obtain the first evidence for the suppressed $B\to DK$ signal
with a significance of $3.2\sigma$. In addition, we report the first
measurements of $A_{DK}$, $R_{D\pi}$ and $A_{D\pi}$. The $R_{DK}$
and $A_{DK}$ results obtained can be used to constrain the UT angle
$\phi_3$ using the ADS method \cite{soni}.

We thank the KEKB group for the excellent operation of the
accelerator; the KEK cryogenics group for the efficient operation of
the solenoid; and the KEK computer group, the National Institute of
Informatics, and the PNNL/EMSL computing group for valuable
computing and SINET4 network support.  We acknowledge support from
the Ministry of Education, Culture, Sports, Science, and Technology
(MEXT) of Japan, the Japan Society for the Promotion of Science
(JSPS), and the Tau-Lepton Physics Research Center of Nagoya
University; the Australian Research Council and the Australian
Department of Industry, Innovation, Science and Research; Austrian
Science Fund under Grant No. P 22742-N16; the National Natural
Science Foundation of China under contract No.~10575109, 10775142,
10825524, 10875115, 10935008 and 11175187; the Ministry of
Education, Youth and Sports of the Czech Republic under contract
No.~MSM0021620859; the Carl Zeiss Foundation, the Deutsche
Forschungsgemeinschaft and the VolkswagenStiftung; the Department of
Science and Technology of India; the Istituto Nazionale di Fisica
Nucleare of Italy; The WCU program of the Ministry Education Science
and Technology, National Research Foundation of Korea Grant No.
2011-0029457, 2012-0008143, 2012R1A1A2008330, 2013R1A1A3007772, BRL
program under NRF Grant No. KRF-2011-0020333, BK21 Plus program, and
GSDC of the Korea Institute of Science and Technology Information;
the Polish Ministry of Science and Higher Education and the National
Science Center; the Ministry of Education and Science of the Russian
Federation and the Russian Federal Agency for Atomic Energy; the
Slovenian Research Agency; the Basque Foundation for Science
(IKERBASQUE) and the UPV/EHU under program UFI 11/55; the Swiss
National Science Foundation; the National Science Council and the
Ministry of Education of Taiwan; and the U.S.\ Department of Energy
and the National Science Foundation. This work is supported by a
Grant-in-Aid from MEXT for Science Research in a Priority Area
(``New Development of Flavor Physics''), and from JSPS for Creative
Scientific Research (``Evolution of Tau-lepton Physics'').


\begin{thebibliography}{99}
\bibitem{ckm} {N. Cabibbo}, Phys. Rev. Lett. {\bf 10}, 531 (1963); {M. Kobayashi and T. Maskawa}, Prog. Theor. Phys. {\bf 49}, 652
(1973).
\bibitem{icc} Thoughout this paper, the addition of the charge conjugate decay mode is implicit unless stated otherwise.
\bibitem {glw} {M. Gronau and D. London, Phys. Lett. B {\bf 253}, 483
(1991); M. Gronau and D. Wyler, Phys. Lett. B {\bf 265}, 172
(1991).}
\bibitem{soni}  { D. Atwood, I. Dunietz and A. Soni}, Phys. Rev. D {\bf 63}, 036005 (2001).

\bibitem{giri} {A. Giri, Yu. Grossman, A. Soffer, and J. Zupan, Phys.
Rev. D {\bf 68}, 054018 (2003).}
\bibitem{BABARADS} P. del Amo Sanchez {\it et al.} (BaBar Collaboration), Phys. Rev. D {\bf 82}, 072006 (2010).
\bibitem{BelleADS} Y. Horii {\it et al.} (Belle Collaboration),
Phys. Rev. Lett. {\bf 106}, 231803 (2011).
\bibitem{CDFADS} T. Aaltonen {\it et al.} (CDF Collaboration), Phys. Rev. D {\bf 84}, 091504(R)
(2011).
\bibitem{LHCbADS} R. Aaij {\it et al.} (LHCb Collaboration), Phys. Lett. B {\bf 712}, 203 (2012).
\bibitem{Negishi} K. Negishi {\it et al.} (Belle Collaboration), Phys. Rev. D {\bf 86}, 011101 (R) (2012).
\bibitem{pdg} J. Beringer {\it et al.} (Particle Data Group), Phys. Rev. D {\bf 86}, 010001 (2012).
\bibitem{AS} D.~Atwood and A.~Soni, Phys. Rev. D \textbf{68}, 033003 (2003).
\bibitem{cleo} N. Lowrey {\it et al.} (CLEO Collaboration), Phys. Rev. D {\bf 80}, 031105(R) (2009).
\bibitem{babar1} J.~P.~Lees {\it et al.} (BaBar Collaboration), Phys. Rev. D {\bf 84}, 012002 (2011).
\bibitem{rama} {Matteo Rama}, arXiv:1307.4384 [hep-ex].
\bibitem{detector}A. Abashian {\it et al.} (Belle Collaboration), Nucl. Instrum. Methods Phys. Res., Sect. A {\bf479},
117 (2002); also, see the detector section in J. Brodzicka {\it et
al.}, Prog. Theor. Exp. Phys., 04D001 (2012).
\bibitem{collider}S. Kurokawa and E. Kikutani, Nucl. Instrum. Methods Phys. Res., Sect. A {\bf499}, 1
(2003), and other papers included in this volume; T. Abe {\it et
al.}, Prog. Theor. Exp. Phys., 03A001 (2013) and following articles
up to 03A011.
\bibitem{pid}E.~Nakano, Nucl. Instrum. Methods Phys. Res. Sect. A {\bf 494}, 402 (2002).
\bibitem{NN} M. Feindt and U. Kerzel, Nucl. Instrum. Methods Phys. Res., Sect. A {\bf 559}, 190 (2006).
\bibitem{sfw} The Fox-Wolfram moments were introduced in G. C. Fox and S. Wolfram,
Phys. Rev. Lett. {\bf 41}, 1581 (1978); the modified moments used in
this paper are described in S. H. Lee {\it et al.} (Belle
Collaboration), Phys. Rev. Lett. {\bf91}, 261801 (2003).
\bibitem{ftag} H. Kakuno {\it et al.} (Belle Collaboration), Nucl. Instrum. Methods
Phys. Res., Sect. A {\bf 533}, 516 (2004).
\end{thebibliography}
\end{document}